\documentclass[a4paper]{article}
\usepackage[cp866]{inputenc}
\usepackage{graphicx}
\date{}
\author{N.\,V.\,\,Ustinov\thanks{e-mail address: n\_ustinov@mail.ru}
\medskip\\
Quantum Field Theory Department, Tomsk State University,\\
36 Lenin Avenue, Tomsk, 634050, Russia
\vspace{-4ex}}
\title{\bf Breather-like pulses in a medium with the permanent dipole moment}
\date{ }
\begin{document}
\maketitle
\begin{abstract}
\noindent 
The solutions of the reduced Maxwell--Bloch equations for an anisotropic 
two-level medium, which describe the propagation of electromagnetic pulses 
having a duration from a few field oscillations, are studied. 
An influence of the permanent dipole moment of the quantum transition on 
dynamics of the pulses and their spectrum is considered.
\end{abstract}


\section{Introduction}

The significant attention was paid during the last years on investigation of 
the nonlinear coherent phenomena in the anisotropic media$\mbox{}^{1-10}$. 
Distinctive feature of such the media is that the stationary states of quantum 
particles being contained in them possess no the parity. 
For this reason, the diagonal elements of the matrix of the dipole moment 
operator and their difference commonly referred to as permanent dipole moment 
(PDM) of the transition are distinct from zero. 

The study of the formation and dynamics of electromagnetic pulses in the 
anisotropic media has great practical importance. 
Modern technologies give an opportunity to produce the semiconductor 
crystals with properties varying over a wide range. 
For example, an anisotropic medium, in which the modulus of the ratio between 
PDM and corresponding off-diagonal matrix element of the dipole moment 
operator lies for different quantum transitions within the limits from 0.15 up 
to 7.1, is resulted in work$\mbox{}^1$. 
Besides, one of the persistent trends in developing the laser physics is to 
obtain the most shorter pulses$\mbox{}^{11,12}$.

The role of PDM in the second harmonic generation in asymmetric semiconductor 
quantum wells was recognized and extensively studied (see, e.g.,$\mbox{}^1$ 
and references therein). 
Its relative influence on the conversion efficiency increases for larger 
pump/harmonic wavelengths, especially at higher pump powers. 
Also, propagation in the resonant optically uniaxial media of the 
two-component electromagnetic pulses consisting of the short-wavelength 
ordinary and long-wavelength extraordinary components was investigated 
in$\mbox{}^{6-9}$. 
It was revealed that, in the conditions of strong interaction between the 
components, pulses can pass through the anisotropic medium in the modes 
distinct from the self-induced transparency. 

The propagation of one-component extremely short electromagnetic pulses of 
femtosecond durations through the media having PDM was considered in 
works$\mbox{}^{2-5,10}$. 
As opposite to the ultrashort pulses, a high-frequency carrier wave is absent 
in the Fourier spectrum of such the pulses. 
It was noted in$\mbox{}^{2}$ that nonzero PDM may have a strong effect on the 
medium response for very short pulses, while an application of the usual 
scheme of the rotating-wave approximation eliminates its influence in the 
longer pulses case. 
Complete integrability of the reduced \mbox{Maxwell--Bloch} equations for the 
two-level medium with PDM in the frameworks of the inverse scattering 
transformation method$\mbox{}^{13-15}$ was established in paper$\mbox{}^{3}$. 
The pulse solutions of these equations on constant (but not arbitrary) 
background were obtained. 
The background was chosen in such a manner that the problem considered was 
reduced to the isotropic medium case. 
Dynamics of these pulses at the pump presence was considered in$\mbox{}^{4}$. 
The system of the full Maxwell--Bloch equations was studied in 
work$\mbox{}^5$, where stationary extremely short pulses decaying 
exponentially and algebraically were found. 
An asymmetry caused by PDM on the polarity of these pulses was revealed. 
Also, the effects of passing the extremely short pulses with a duration up to 
several oscillations of electromagnetic field through a medium possessing PDM 
were investigated numerically$\mbox{}^{10}$. 
It was shown that there exists solitary stable bipolar signal with a nonzero 
time area (nonzero breather).
The recent results summarized above make valuable to clarify the role of PDM 
in the formation the one-component pulses of the self-induced transparency 
(including, the extremely short pulses and ultrashort ones as the limiting 
cases). 
In present report, we shall study the exact solutions of the breather type of 
system of the reduced Maxwell--Bloch equations for an anisotropic medium. 

\section{The model and basic equations}

Let's consider optically uniaxial medium, whose anisotropy is created by the 
electric field. 
It splits the energy levels owing to the Stark effect, retaining the 
degeneration of electronic levels on the absolute value of projection $M$ of 
the total angular momentum. 
Thus, $\pi$-transition ($\Delta M=0$) and doubly degenerate 
$\sigma$-transitions ($|\Delta M|=1$) are formed in an electronic subsystem. 

Let the electromagnetic pulse propagate in the positive direction of $y$ axis 
perpendicular to the optical axis $z$ of the medium. 
Suppose that only the extraordinary component $E_e$ of its electric field, 
which is parallel to $z$ axis, is distinct from zero. 
One can show that such the pulse will interact with the $\pi$-transition only. 
The system of reduced Maxwell--Bloch equations describing this process in the 
unidirectional propagation approximation$\mbox{}^{16}$ has next form: 
\begin{equation}
\frac{\partial\sigma _3}{\partial t}=i\frac{d}{\hbar}E_e(\sigma-\sigma^*),
\label{1}
\end{equation}
\begin{equation}
\frac{\partial\sigma}{\partial t}=i\Bigl(\omega_0+\frac{D}{\hbar}E_e\Bigr)
\sigma+2i\frac{d}{\hbar}E_e\sigma_3,
\label{2}
\end{equation}
\begin{equation}
\frac{\partial E_e}{\partial y}+\frac{n_e}{c}\frac{\partial E_e}{\partial t}=
-2\pi i\frac{Nd\,\omega_0}{n_ec}(\sigma-\sigma^*), 
\label{3}
\end{equation}
where  $\sigma_3=(\rho_{22}-\rho_{11})/2$ is the population inversion of the 
quantum level; $\sigma=\rho_{12}$; $\rho_{jk}$ ($k=1,2$) are the elements of 
the density matrix; $d$, $D$ and $\omega_0$ are the dipole moment, PDM and the 
frequency of $\pi$-transition, respectively; $n_e$ is the extraordinary 
refractive index; $N$ is the concentration of the $\pi$-transitions. 
In the equations given, we ignore the relaxation and inhomogeneous broadening 
of the spectral line of resonant absorption. 

System (\ref{1})--(\ref{3}) coincides with the reduced Maxwell--Bloch 
equations for the isotropic media$\mbox{}^{16}$ if $D=0$. 
One can see that the extraordinary component of electric field fulfills here 
two functions: it causes the quantum transitions and, simultaneously, shifts 
dynamically their frequency. 
In the case of the two-component pulses$\mbox{}^{6-9}$, these functions were 
executed by its short-wavelength ordinary and long-wavelength extraordinary 
components accordingly. 
In so doing, the ordinary component generated extraordinary one due to PDM of 
the resonant $\sigma$-transitions. 

It is convenient for subsequent consideration to introduce new variables
\begin{equation}
u=\frac{dE_e}{\hbar\omega_0},\quad\tau=\omega_0\Bigl(t-\frac{n_e}{c}y\Bigr),
\quad\eta=2\,\frac{\pi N d^{\,2}}{n_ec\hbar}\,y. 
\label{v}
\end{equation}
Then, equations (\ref{1})--(\ref{3}) are rewritten as 
\begin{equation}
\frac{\partial\sigma_3}{\partial\tau}=iu(\sigma-\sigma^*),
\label{rmb1}
\end{equation}
\begin{equation}
\frac{\partial\sigma}{\partial\tau}=i(1+2ku)\sigma+2iu\sigma_3,
\label{rmb2}
\end{equation}
\begin{equation}
\frac{\partial u}{\partial\eta}=i(\sigma^*-\sigma), 
\label{rmb3}
\end{equation}
where 
$$
k=\frac{D}{2d}\,. 
$$

An integrability of equations (\ref{rmb1})--(\ref{rmb3}) with arbitrary 
parameter $k$ in the frameworks of the inverse scattering transformation 
method$\mbox{}^{13-15}$ was revealed in$\mbox{}^3$. 
The corresponding overdetermined system of linear equations (Lax pair)
has the next form
\begin{equation}
\left\{
\begin{array}{l}
\displaystyle\frac{\partial\psi}{\partial\tau_{\mathstrut}}=
L(\lambda)\psi(\lambda),\\ 
\displaystyle\frac{\partial\psi^{\mathstrut}}{\partial\eta}=
A(\lambda)\psi(\lambda),
\end{array}
\right.
\label{Lax}
\end{equation}
where $\psi=\psi(\tau,\eta,\lambda)=(\psi_1,\psi_2)^T$ is the vector solution 
of the Lax pair, $\lambda$ is so-called spectral parameter, matrices 
$L(\lambda)$ and $A(\lambda)$ are defined as follows
$$
L(\lambda)=\left( 
\begin{array}{cc}
\displaystyle\frac{ik}{2\sqrt{1+k^2}}+i\sqrt{1+k^2}u&
\displaystyle\frac{i\lambda}{2}\\     
\displaystyle\frac{i\lambda}{2}&
\displaystyle-\frac{ik}{2\sqrt{1+k^2}}-i\sqrt{1+k^2}u
\end{array}
\right),
\label{L}
$$
$$
A(\lambda)=\displaystyle\frac{i(1+k^2)\lambda}
{1-(1+k^2)\lambda^2_{\mathstrut}}\left( 
\begin{array}{cc}
\displaystyle\frac{2k\sigma_3-\sigma-\sigma^*}
{\sqrt{1+k^2}\,\lambda_{\mathstrut}}&
\displaystyle2\sigma_3+k(\sigma+\sigma^*)+\sqrt{1+k^2}(\sigma-\sigma^*)\\     
\displaystyle2\sigma_3+k(\sigma+\sigma^*)+\sqrt{1+k^2}(\sigma^*-\sigma)&
\displaystyle\frac{-2k\sigma_3+\sigma+{\sigma^*}^{\mathstrut}}
{\sqrt{1+k^2}\,\lambda}
\end{array}
\right).
\label{A}
$$
One can check by direct calculation that the compatibility condition of both 
the equations of Lax pair (\ref{Lax}):
\begin{equation}
\frac{\partial L(\lambda)}{\partial\eta}-\frac{\partial A(\lambda)}{\partial\tau}
+[L(\lambda),A(\lambda)]=0,
\label{cc}
\end{equation}
is equivalent to system (\ref{rmb1})--(\ref{rmb3}).

\section{Exponentially and rationally decreasing breather-like pulses}

Being integrable by the inverse scattering transformation method, equations 
(\ref{rmb1})--(\ref{rmb3}) have the multisoliton solutions. 
The one-soliton pulse and its algebraic limit can be obtained by direct 
integration of these equations in stationary case as it was done, for example, 
for the system of the full Maxwell--Bloch equations$\mbox{}^5$. 
The formulas arising at it differ from ones presented in$\mbox{}^5$ by the 
definition of the pulse velocity only. 
Applying the Darboux transformation technique$\mbox{}^{17}$ to find the 
two-soliton breather-like pulse solution of system (\ref{rmb1})--(\ref{rmb3}) 
on the zero background, we come to the following expression for variable $u$: 
\begin{equation}
u=-\frac{2}{\sqrt{1+k^2}}\,\frac{\partial}{\partial\tau}\,\,
\mbox{arctan}\left[\frac{4ka_I\Omega\sinh A_R+Z_+T^{-1}\cos A_I}
{Z_-\Omega\cosh A_R-4ka_RT^{-1}\sin A_I}\right]  
\label{u_br}
\end{equation}
where
$$
A_R=\frac{T^{-1}}{\sqrt{1+k^2}}\Bigl(\tau+16(1+k^2)(Z_+^2+4Z+16k^2a_R^2)
\frac{Z\sigma_0}{A}\eta\Bigr)+c_1,
$$
$$
A_I=\frac{\Omega}{\sqrt{1+k^2}}\Bigl(\tau-16(1+k^2)(Z_-^2-4Z-16k^2a_I^2)
\frac{Z\sigma_0}{A}\eta\Bigr)+c_2,
$$
$$
A=Z_+^4+8Z_+^2(Z-4k^2a_R^2)+16(1-4a_R^2)(Z^2-4k^4a_R^2), 
$$
\begin{equation}
\Omega=a_RZ_+/Z,\quad T^{-1}=a_IZ_-/Z,\quad Z=4(a_R^2+a_I^2),\quad 
Z_{\pm}=Z\pm k^2. 
\label{Omega_T}
\end{equation}
Real constants $a_R$, $a_I$, $c_1$ and $c_2$ are the free parameters of the 
pulse. 
Constant $\sigma_0$ is an initial population of the quantum level 
($|\sigma_0|\le1/2$). 
The formulas for variables $\sigma_3$ and $\sigma$ are cumbersome and omitted. 

The time area of this two-soliton solution is distinct from zero:
$$
\int\limits_{-\infty}^{\infty}\!\!{u}\,d\tau=-
\frac{4\,\mbox{sign}\,(k)}{\sqrt{1+k^2}}\,\,
\mbox{arctan}\left[|Tk|\,\frac{a_I^2}{a_R^2+a_I^2}\right].
$$ 
For this reason, we refer to it as the breather-like pulse to distinguish from 
the well-known breather pulse in isotropic media, whose time area is equal to 
zero. 
An existence of the breather type solutions with nonzero area was established 
numerically for system (\ref{rmb1})--(\ref{rmb3}) in$\mbox{}^{10}$. 

Quantities $\Omega$ and $T$, which characterize the frequency and duration of 
the pulse, can be regarded as its free parameters instead of $a_R$ and 
$a_I$. 
Then, one finds from equations (\ref{Omega_T})
$$
a_R=\frac{\Omega}{2}\left(1+\frac{R_+}{\Omega T}\right),
$$
$$
a_I=\frac{1+R_-}{2T}, 
$$
where
$$
R_\pm=\frac{1}{\sqrt{2}\,}\sqrt{r\pm((\Omega^2-k^2)T^2-1)}\,,
$$
$$
r=\sqrt{((\Omega^2-k^2)T^2+1)^2+4k^2T^2}\,.
$$
Here the arithmetic root is taken in the definition of $r$, and the signs of 
$R_+$ and $R_-$ are chosen in such a manner that $R_+R_-=\Omega T$. 
Without loss of a generality, we shall suppose in the sequel that parameters 
$a_R$ and $a_I$ (or $\Omega$ and $T$) are positive. 
If $\Omega T\gg1$, then solution (\ref{u_br}) is nothing but the pulse with 
higher-frequency filling (ultrashort pulse), while at $\Omega T<1$ variable 
$u$ may no change the polarity as it happens for the extremely short pulses. 

Expanding the right-hand side of equation (\ref{u_br}) in the Taylor series at 
a neighborhood of point $k=0$ and retaining the first two terms, we come to 
formulas
$$
u=u_0+ku_1,
$$
\begin{equation}
u_0=4\Omega\,\frac{\Omega T\cosh B_R\sin B_I+\sinh B_R\cos B_I}
{\Omega^2T^2(\cosh 2B_R+1)+\cos 2B_I+1},
\label{u_0}
\end{equation}
\begin{equation}
u_1=-\,\frac{4\Omega^2 T^2(\cosh2B_R+1)(\cos2B_I+1)}
{(\Omega^2T^2(\cosh2B_R+1)+\cos2B_I+1)^2},
\label{u_1}
\end{equation}
where 
$$
B_R=\frac{\tau}{T}+\frac{4T((\Omega^2+1)T^2+1)\sigma_0\eta}
{((\Omega+1)^2T^2+1)((\Omega-1)^2T^2+1)},
$$
$$
B_I=\Omega\tau-\frac{4\Omega T^2((\Omega^2-1)T^2+1)\sigma_0\eta}
{((\Omega+1)^2T^2+1)((\Omega-1)^2T^2+1)}.
$$
It follows from the expressions presented that the modulus of the Fourier 
transform of $u_0$ achieves the maxima on odd harmonics of the basic frequency 
$\Omega$, while the modulus of the Fourier transform of $u_1$ has them on even 
harmonics, including zeroth one. 
Since the generation of the secondary harmonics due to PDM is nonlinear effect, 
they are more localized at the center of the pulse. 
In the case of breather-like pulses, an asymmetry induced by PDM manifests 
itself in that the signs of $k$ and the zero harmonic are opposite. 
Similar asymmetry on the polarity of a signal for the extremely short pulses 
was revealed under analytical and numerical investigations in$\mbox{}^{5,10}$. 

The conclusions made above are most obvious for the pulses with slowly varying 
envelope. 
Indeed, if condition $\Omega T\gg1$ holds, then formulas (\ref{u_0}) and 
(\ref{u_1}) become simpler: 
\begin{equation}
u_0=\frac{2\sin B_I}{T\cosh B_R},
\label{u_0_sve}
\end{equation}
\begin{equation}
u_1=-\left(\frac{2\cos B_I}{\Omega T\cosh B_R}\right)^2\!.
\label{u_1_sve}
\end{equation}
Carrying out the Fourier transform with expressions (\ref{u_0_sve}) and 
(\ref{u_1_sve}): 
$$
F(\nu,u_{0,1})=\int\limits_{-\infty}^{\infty}\mbox{e}^{\displaystyle i\nu \tau}
u_{0,1}\,d\tau,
$$
one obtains 
$$
F(\nu,u_0)=-i\pi\exp(-iT\theta_R\nu)\left(
\frac{\exp i(\theta_I-\Omega T\theta_R)}{\cosh\pi T(\nu+\Omega)/2}-
\frac{\exp i(\Omega T\theta_R-\theta_I)}{\cosh\pi T(\nu-\Omega)/2}\right),
$$
$$
F(\nu,u_1)=-\pi\frac{\exp(-iT\theta_R\nu)}{\Omega^2}
\left(\frac{2\nu}{\sinh\pi T\nu/2}+(\nu+2\Omega)
\frac{\exp2i(\theta_I-\Omega T\theta_R)}{\sinh\pi T(\nu+2\Omega)/2}+
\mbox{}\right.
$$
$$
\left.\mbox{}+(\nu-2\Omega)
\frac{\exp2i(\Omega T\theta_R-\theta_I)}{\sinh\pi T(\nu-2\Omega)/2}\right),
$$
where $\theta_R$ and $\theta_I$ are the values of $B_R$ and $B_I$ at $\tau=0$. 

It is seen that the maxima of the moduli of $F(\nu,u_0)$ and $F(\nu,u_1)$ are 
reached at $\nu=\Omega$ and at $\nu=0,2\Omega$, respectively. 
The width of the spectral lines is equal to $T^{-1}$, while the maximum values 
of $|F(\nu, u_1)|$ are proportional to $\Omega^{-2}T^{-1}$. 
Thus, an efficiency of the secondary harmonic generation grows with reducing 
the carrier frequency of the pulse. 
This agrees with the results of the previous studies (see, e.g.,$\mbox{}^1$). 
An influence of PDM on the pulses having the high-frequency filling is weak. 
This is due to the fact that an average value of the frequency detuning $2ku$ 
(see formula (\ref{rmb2})) on the duration of such the pulse tends to zero. 

Let us return to expression (\ref {u_br}) again. 
The time dependence of $u$ and $\sigma_3$ is plotted below.
\begin{figure}[ht]
\centering
\includegraphics[width=3.34in]{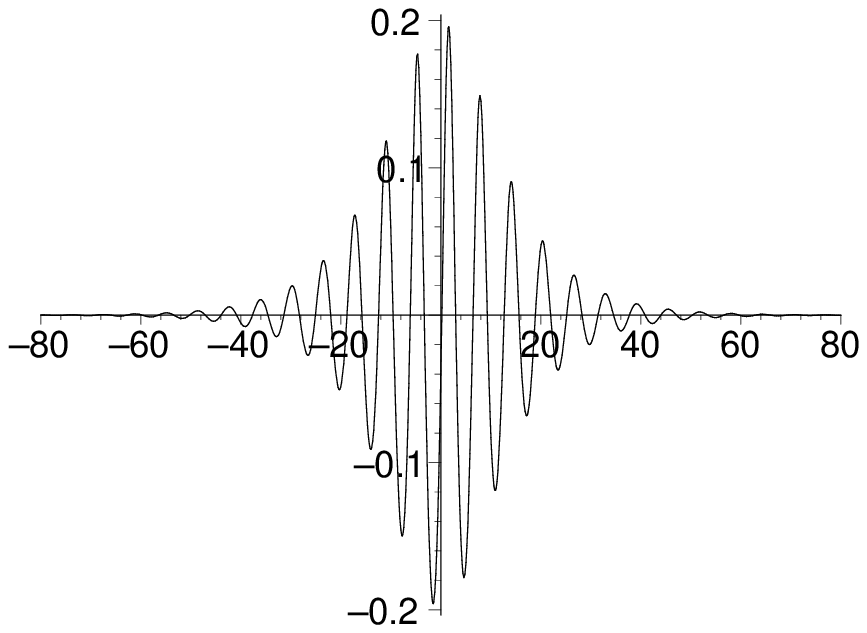}
\includegraphics[width=3.34in]{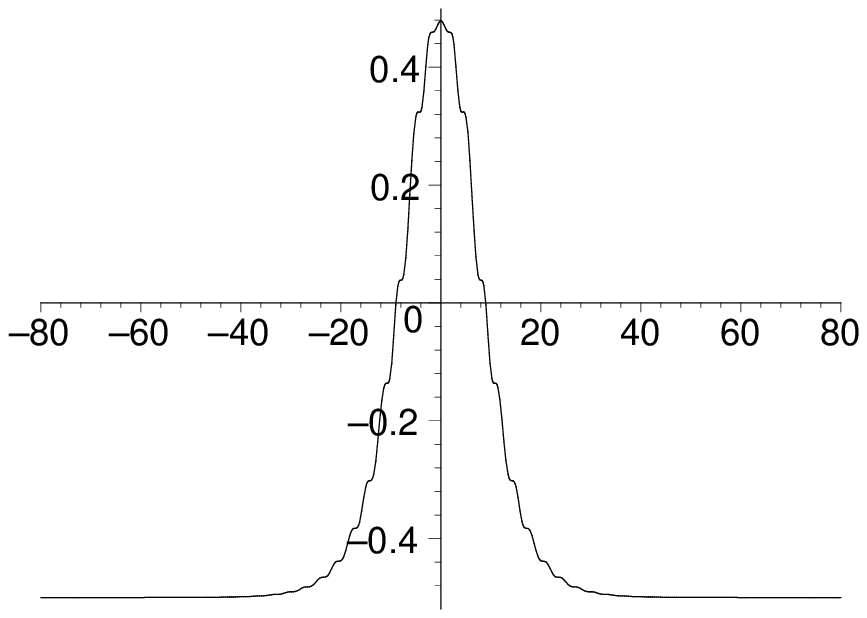}
\begin{picture}(0,0)
\put(-115,170){$u$}
\put(-40,108){$\tau$}
\put(130,170){$\sigma_3$}
\put(205,108){$\tau$}
\put(-215,170){\small\bf a}
\put(30,170){\small\bf b}
\end{picture}
\vskip-0.5cm
\centerline{\small\bf Fig.\,1. \rm Profiles of $u$ and $\sigma_3$ with $k=1$, 
$\sigma_0=-0.5$, $\tilde\Omega=1$ and $\tilde T=10$.}
\end{figure}
Parameters $\tilde\Omega=\Omega/\sqrt{1+k^2}$ and $\tilde T=\sqrt{1+k^2}\,T$ 
of the pulse are selected so that it excites the medium strongly (see 
Fig.~1b). 
The curve of the modulus of the Fourier transform of $F(\nu,u)$ is given on 
Fig. 2. 
\begin{figure}[ht]
\vskip0.0cm
\centering
\includegraphics[width=3.5in]{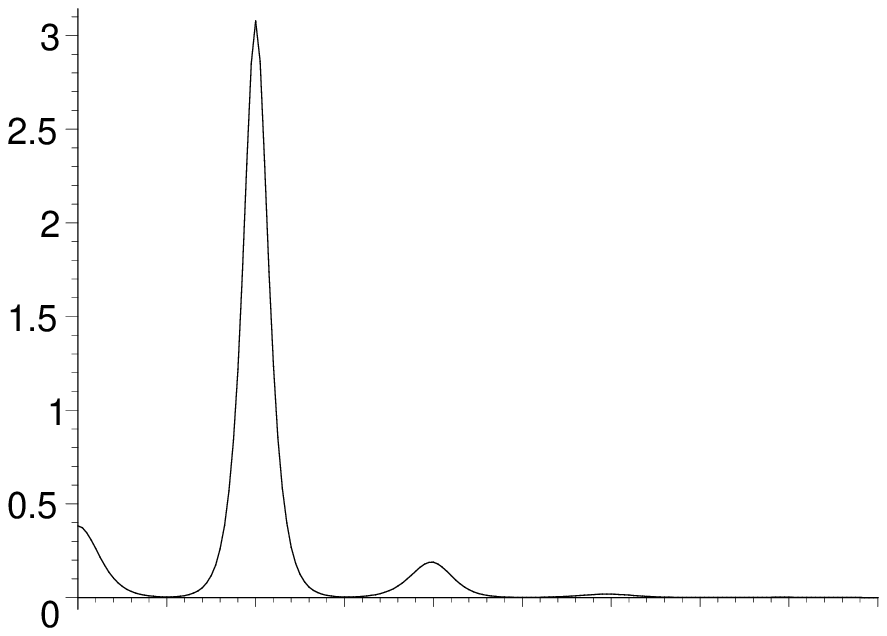}
\begin{picture}(0,0)
\put(-225,170){$|F(\nu,u)|$}
\put(-40,27){$\nu$}
\put(-190,7){$\tilde\Omega$}
\put(-148,7){$2\tilde\Omega$}
\put(-102,7){$3\tilde\Omega$}
\put(-58,7){$4\tilde\Omega$}
\end{picture}
\vskip-0.2cm
\centerline{\small\bf Fig.\,2. \rm Modulus of Fourier transform $F(\nu,u)$ of 
the pulse presented on Fig.\,1.}
\end{figure}
Note, that the zeroth harmonic exists in the Fourier spectrum in accordance 
with that the time area of the breather-like pulses differs from zero. 
For the pulses with $\Omega T\approx1$, this fact was established 
in$\mbox{}^{10}$ (see Fig. 3 also). 
\begin{figure}[ht]
\vskip0.0cm
\centering
\includegraphics[width=3.5in]{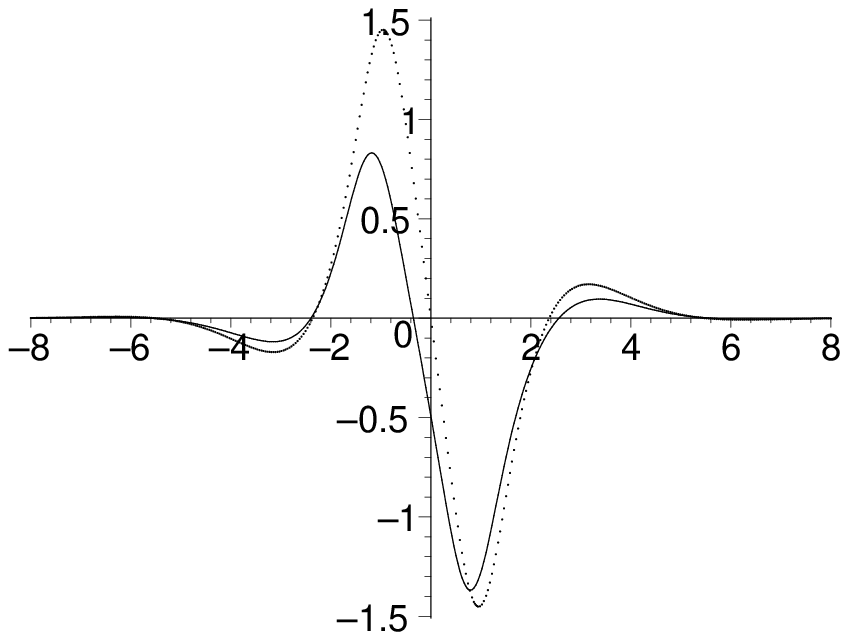}
\begin{picture}(0,0)
\put(-120,170){$u$}
\put(-40,100){$\tau$}
\end{picture}
\vskip-0.5cm
\centerline{\small\bf Fig.\,3. \rm Profiles of $u$ with $k=1$, 
$\sigma_0=-0.5$, $\tilde\Omega=1$, $\tilde T=1$ (solid curve) and with $k=0$ 
(dotted line).}
\end{figure}
Since the principal maximum on Fig.~2 is arranged to the left from 
$\tilde\Omega$, the secondary harmonics are generated most effectively by the 
pulses, whose basic carrier frequency on an input of the anisotropic medium is 
lesser than the resonant frequency. 
The same effect takes place under reducing a duration of incident pulse due to 
broadening of its spectral width. 
Also, it is possible to show that the position of the principal maximum on 
$\nu$ axis in the Fourier spectrum is displaced to a red region with growing 
$|k|$. 
The curve on Fig.~4 shows a difference of the spectral structure of $u$'s with 
identical $\tilde\Omega$ and $\tilde T$ in the cases of anisotropic and 
isotropic media. 
The lapse to the right of $\tilde\Omega$ is a consequence of increasing an 
asymmetry of the principal peak of the Fourier spectrum in the medium with 
PDM. 
\begin{figure}[ht]
\vskip0.0cm
\centering
\includegraphics[width=3.5in]{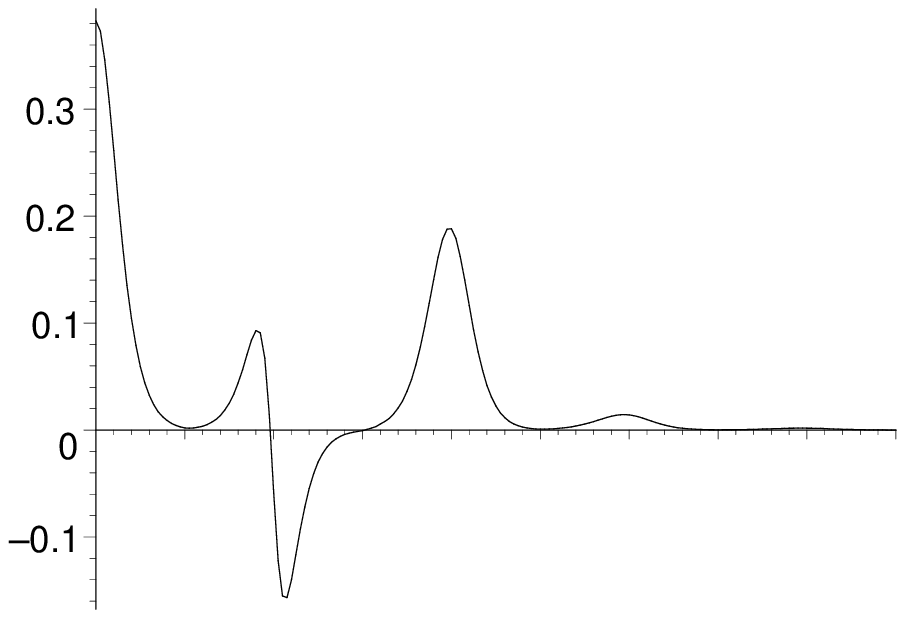}
\begin{picture}(0,0)
\put(-225,170){$\Delta F$}
\put(-40,70){$\nu$}
\put(-190,49){$\tilde\Omega$}
\put(-148,49){$2\tilde\Omega$}
\put(-102,49){$3\tilde\Omega$}
\put(-58,49){$4\tilde\Omega$}
\end{picture}
\vskip-0.5cm
\centerline{
\parbox{14.0cm}{\small\bf Fig.\,4. \rm Difference $\Delta F$ of the moduli of 
Fourier transform $F(\nu,u)$ of the breather-like pulses with $k=1$ and 
$k=0$ (isotropic case). Values of the remaining parameters are the same as 
for the~pulse plotted in Fig.\,1.}}
\end{figure}

In limiting case $2(a_R+ia_I)\to k\exp i\varphi$ (or $T\to\infty$) formula 
(\ref{u_br}) yields 
\begin{equation}
u=-2\,\frac{\Omega\,\cot\varphi}{\sqrt{1+k^2}}\,
\frac{\zeta_1\sin^2\!\varphi\,\cos\zeta_2+(\cos^2\!\varphi-2)\sin\zeta_2+
2\sin\varphi}{(\zeta_1\sin\varphi+\cos\zeta_2)^2+\cot^2\!\varphi\,
(1-\sin\varphi\,\sin\zeta_2)^2}, 
\label{u_r}
\end{equation}
where  
$$
\zeta_1=\Omega\tau+\frac{4\Omega(1+\Omega^2)\sigma_0\eta}{(1-\Omega^2)^2}+c_1,
\quad\zeta_2=\Omega\tau+\frac{4\Omega\sigma_0\eta}{1-\Omega^2}+c_2, 
$$
$$
\Omega=\frac{k\,\cos\varphi}{\sqrt{1+k^2}}.
$$
Here variable $u$ decreases rationally. 
An arbitrary parameter of the pulse is real constant $\varphi$ determining 
its carrier frequency, which is always less resonant one. 

Existence of rationally decreasing pulses is the distinctive feature of the 
anisotropic media. 
So, algebraic one-component extremely short pulse was found in$\mbox{}^5$, and 
two-component one-parameter pulses were constructed in$\mbox{}^9$. 
Curves of $u$, $\sigma_3$ and $|F(\nu,u)|$ for the pulse (\ref{u_r}) are 
represented in Figs~5 and 6. 
\begin{figure}[ht]
\centering
\includegraphics[width=3.34in]{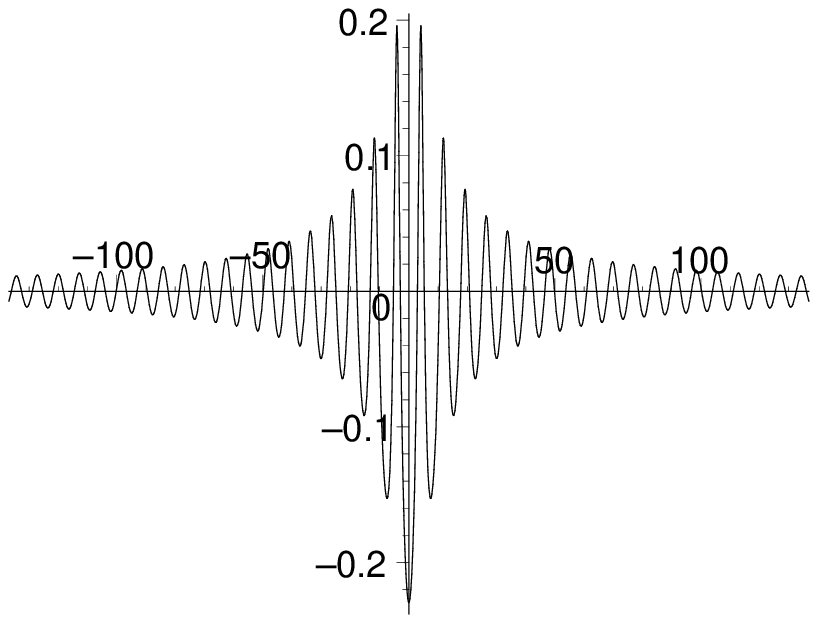}
\includegraphics[width=3.34in]{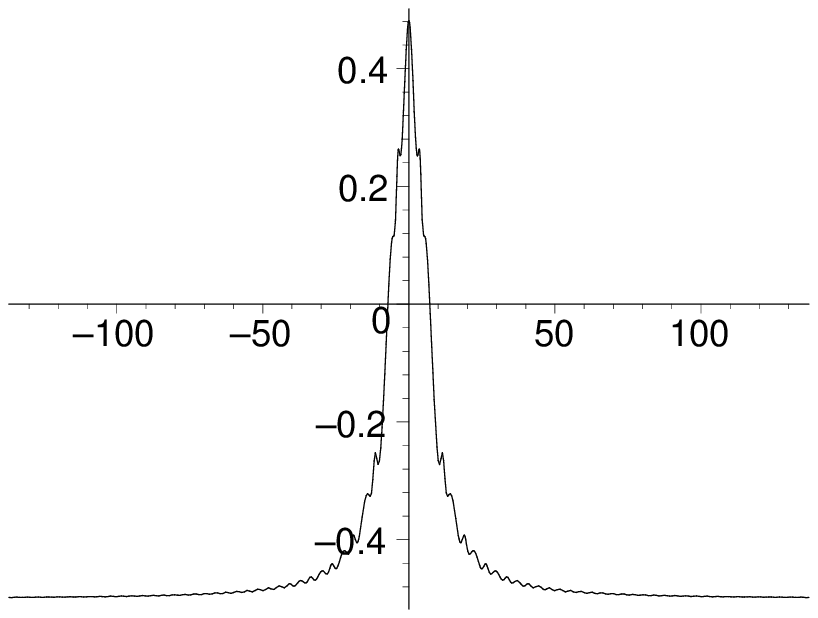}
\begin{picture}(0,0)
\put(-115,170){$u$}
\put(-36,115){$\tau$}
\put(130,170){$\sigma_3$}
\put(205,108){$\tau$}
\put(-215,170){\small\bf a}
\put(30,170){\small\bf b}
\end{picture}
\vskip-0.5cm
\centerline{
\parbox{14.0cm}{\small\bf Fig.\,5. \rm Profiles of $u$ and $\sigma_3$ of the 
rationally decreasing breather-like pulse with $k=3$, $\sigma_0=-0.5$ and 
$\varphi=\pi/8$ ($\Omega\approx0.88$).}}
\end{figure}
Comparing curves on Figs~2 and 6, one can see that the position on $\nu$ axis 
of maxima of the secondary harmonics is more shifted in the red region for 
the rationally decreasing pulses. 
Besides, the expressed asymmetry has not only the principal peak, but the 
secondary peaks of the Fourier spectrum also. 
\begin{figure}[ht]
\vskip0.0cm
\centering
\includegraphics[width=3.5in]{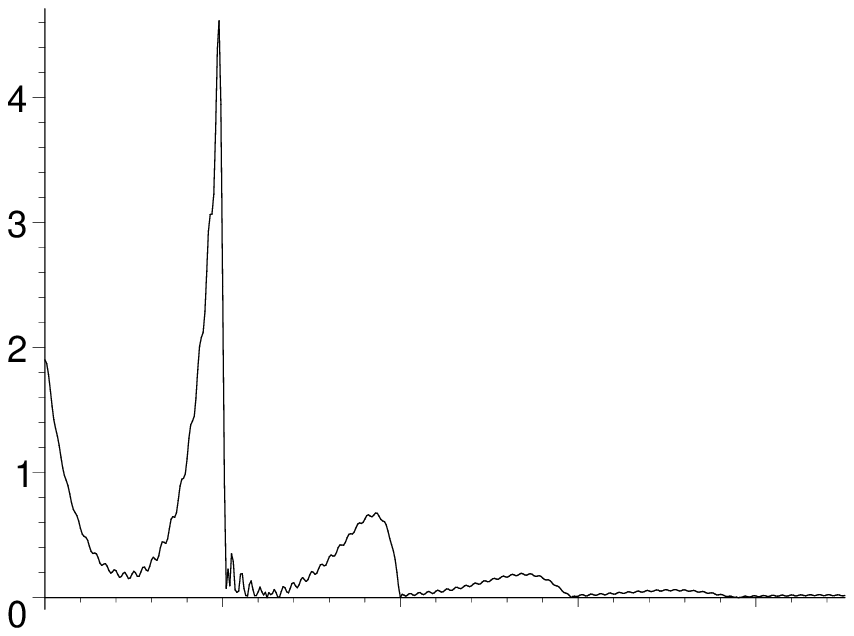}
\begin{picture}(0,0)
\put(-225,170){$|F(\nu,u)|$}
\put(-40,29){$\nu$}
\put(-190,7){$\Omega$}
\put(-148,7){$2\Omega$}
\put(-102,7){$3\Omega$}
\put(-58,7){$4\Omega$}
\end{picture}
\vskip-0.0cm
\centerline{
\parbox{14.0cm}{\small\bf Fig.\,6. \rm Modulus of Fourier transform 
$F(\nu,u)$ of the rationally decreasing breather-like pulse. 
Values of the parameters correspond to the pulse on Fig.\,5.}}
\vskip-0.2cm
\end{figure}

Performing limiting procedures $\varphi\to\pm\pi/2$ and $-c_1\to\pm\cos c_2$, 
we obtain from equation (\ref{u_r}):
\begin{equation}
u=-\frac{2k}{1+k^2+k^2(\tau+4\sigma_0\eta)^2}.
\label{u_r1}
\end{equation}
If we impose additional condition $c_2\to\pm\pi/2$ in so doing, then, instead 
of (\ref{u_r1}), formula (\ref{u_r}) turns into following
\begin{equation}
u=-6k\,\frac{w_1}{w_2},
\label{u_r2}
\end{equation}
where 
$$
w_1=k^4(\tau+4\sigma_0\eta)^4+6k^2(\tau+4\sigma_0\eta)
\Bigl((1+k^2)\tau+4(1-3k^2)\sigma_0\eta\Bigr)-3(1+k^2)^2,
$$
$$
\begin{array}{c}
w_2=k^6(\tau+4\sigma_0\eta)^6+3k^4(\tau+4\sigma_0\eta)^3
\Bigl((1+k^2)\tau+4(1+9k^2)\sigma_0\eta\Bigr)_{\mathstrut}+\mbox{}\\
\mbox{}+9k^2\Bigl(3(1+k^2)^2\tau^2+8(1+k^2)(3-k^2)\sigma_0\tau\eta+
16(3-2k^2+11k^4)\sigma_0^2\eta^2\Bigr)^{\mathstrut}+9(1+k^2)^3.
\end{array}
$$

At the end of this section, we discuss briefly the spectral 
data$\mbox{}^{13-15}$ corresponding to the solutions studied. 
The discrete spectrum of breather-like pulse (\ref{u_br}) includes four simple 
points. 
They are located on the plane of the Lax pair (\ref{Lax}) spectral parameter 
at 
$$
\lambda=\pm\frac{4(a_R+ia_I)^2-k^2}{2(a_R+ia_I)\sqrt{1+k^2}},\quad
\lambda=\pm\frac{4(a_R-ia_I)^2-k^2}{2(a_R-ia_I)\sqrt{1+k^2}}.
$$
The points of the discrete spectrum of rationally decreasing pulse (\ref{u_r}) 
are doubly degenerate and located at 
$$
\lambda=\pm\frac{ik\sin\varphi}{\sqrt{1+k^2}}.
$$
Solution (\ref{u_r1}) has two simple points of the discrete spectrum at 
$\lambda=\pm ik/\sqrt{1+k^2}$. 
This rational pulse is a particular case of the one-soliton solution of system 
(\ref{rmb1})--(\ref{rmb3}), which is stationary. 
The points of the discrete spectrum of solution (\ref{u_r2}) are the same as 
for previous one, but they are doubly degenerate. 

\section{Conclusion}
In the present report, we have considered the propagation through the 
two-level medium possessing PDM of the electromagnetic pulses with duration 
from a several oscillations of the field. 
The explicit breather-like solutions of corresponding system of the reduced 
Maxwell--Bloch equations are studied. 
Unlike to the breathers in an isotropic medium, these ones have the time area 
distinct from zero. 
An existence of the solutions of such a kind (namely, the nonzero breather) 
was revealed under numerical analysis in$\mbox{}^{10}$.
As in the case of the extremely short electromagnetic pulses$\mbox{}^{5,10}$, 
it is shown here that an asymmetry on the polarity of a signal takes place for 
the breather-like pulses of arbitrary duration: the signs of the zero harmonic 
and PDM are opposite. 
It follows from the Fourier analysis of the solutions under discussion that an 
influence of PDM on an efficiency of the secondary harmonics generation grows 
with a reduction of the basic carrier frequency of the pulse and its duration 
shortening. 
As a result, the pulses, whose carrier frequency on an input of the medium is 
less than resonant one, will generate the secondary harmonics most 
effectively. 
This concern, in particular, the rationally decreasing breather-like pulses 
that exist only in the case of the anisotropic media. 

\section{Acknowledgment}

This work was supported by the RBRF grant 05--02--16422\,a.

\end{document}